\documentclass[twocolumn]{aa}

\bibpunct{(}{)}{;}{a}{}{,}

\usepackage{graphicx}
\usepackage[varg]{txfonts}
\def\ks{km s$^{-1}$}
\def\d{$^\circ$}
\def\m{$^\prime$}
\def\s{$^{\prime\prime}$}
\def\hh{$^{\mathrm h}$}
\def\mm{$^{\mathrm m}$}
\def\ss{$^{\mathrm s}$}
\def\cm3{cm$^{-3}$}

\def\2{$^{12}$CO}
\def\3{$^{13}$CO}
\def\H{HCO$^{+}$}
\def\msol{M$_\odot$}
\begin{document}

\title{ASTE observations in the 345 GHz window towards the HII region N113 of the Large Magellanic Cloud} 
\subtitle{}

\author{S. Paron \inst{1,2}
\and M. E. Ortega \inst{1}
\and M. Cunningham \inst{3}
\and P. A. Jones \inst{3}
\and M. Rubio \inst{4}
\and C. Fari\~{n}a \inst{5}
\and S. Komugi \inst{6}
}

\institute{Instituto de Astronom\'{\i}a y F\'{\i}sica del Espacio (IAFE, CONICET-UBA),
             CC 67, Suc. 28, 1428 Buenos Aires, Argentina \\
            \email{sparon@iafe.uba.ar}
\and FADU and CBC, Universidad de Buenos Aires, Argentina 
\and School of Physics, University of New South Wales, Sydney, NSW 2052, Australia  
\and Departamento de Astronom\'{\i}a, Universidad de Chile, Casilla 36-D, Santiago, Chile 
\and Isaac Newton Group of Telescopes, E38700, La Palma, Spain
\and National Astronomical Observatory of Japan, 2-21-1 Osawa, Mitaka, Tokyo 181-8588
}

\offprints{S. Paron}

   \date{Received <date>; Accepted <date>}

\abstract{}{N113 is an HII region located in the central part of the Large Magellanic Cloud (LMC) with an associated 
molecular cloud very rich in molecular species. Most of the previously observed molecular lines cover the frequency 
range 85--270 GHz. Thus, a survey and study of lines at the 345 GHz window is required in order to have a
more complete understanding of the chemistry and excitation conditions of the region.}{We mapped a region of 2\farcm5 $\times$ 2\farcm5~centered 
at N113 using the Atacama Submillimeter Telescope Experiment in the \3 J=3--2 line with an angular and spectral resolution
of 22\s~and 0.11 \ks, respectively. In addition, we observed 16 molecular lines as single pointings towards its center.}
{For the molecular cloud associated with N113, from the \3 J=3--2 map we estimate LTE and virial masses of about $1\times10^{4}$~and 
$4.5\times10^{4}$ \msol, respectively. Additionally, from the dust continuum emission at 500 $\mu$m we obtain a mass
of gas of $7\times10^3$ \msol. 
Towards the cloud center we detected emission from: \2, \3, C$^{18}$O (3--2), HCN, HNC, \H, C$_{2}$H (4--3), and CS (7--6);
being the first reported detection of HCN, HNC, and C$_{2}$H (4--3) lines from this region. 
The CS (7--6) which was previously tentatively detected is confirmed in this study. 
By analyzing the HCN, HNC, and C$_{2}$H, we suggest that their emission may arise from a photodissociation region (PDR).
Moreover, we suggest that the chemistry involving the C$_{2}$H in N113 can be similar to that in Galactic PDRs. 
Using the HCN J=4--3, J=3--2, and J=1--0 lines in a RADEX analysis we conclude that we are observing very high density gas, 
between some 10$^{5}$ and 10$^{7}$ cm$^{-3}$.}{}

\keywords{Galaxies: ISM -- (Galaxies:) Magellanic Clouds -- (ISM:) HII regions -- ISM: individual objects: N113 -- ISM: molecules}

\maketitle 

\section{Introduction}

N113 is an HII region located in the central part of the Large Magellanic Cloud (LMC). It hosts two H$_{2}$O masers, being one
of them the most intense of the Magellanic Clouds \citep{whiteoak86,lazen02,imai13}, as well as an OH maser \citep{brooks97}. 
This region is associated with a molecular cloud which is clumpy \citep{seale12} and active in star formation. 
Indeed, three young stellar clusters, NGC 1874, NGC 1876, and
NGC 1877, are related to N113 \citep{bica92} and several young stellar objects (YSOs) have been found embedded in the 
molecular gas associated with this HII region \citep{sewilo10,seale12,carlson12}.

One of the main motivations of molecular observational studies towards different regions in the LMC (e.g. \citealt{joha94}) 
is to study a low metallicity interstellar medium (ISM), whose physical conditions may resemble those that 
existed in the early Milky Way, and thus can shed light on the primeval processes of star formation. 
The N113 associated molecular cloud is one of the richest in the LMC and there are several studies presenting observations, 
with different resolutions and sensitivities, of a large number of molecular lines (\citealt{wong06,wang09}, and references therein). 
However, most of the molecular lines observed towards N113 cover the frequency range 85--270 GHz (see \citealt{chin97,wang09}). 
Only three lines were successfully observed at higher frequencies: \3 and \2 J=3--2 at 330.56 and 345.79 GHz, respectively, 
and \H~J=4--3 at 356.73 GHz. Thus, a survey and study of molecular lines at the 345 GHz window  (324--372 GHz) is required in order to have a 
more complete understanding of the chemistry and excitation conditions of the region. Therefore, we used the Atacama Submillimeter
Telescope Experiment (ASTE) to map N113 in the \3 J=3--2 transition and to observe 16 molecular lines within the 345 GHz window 
towards its center.

\begin{figure*}[tt]
\centering
\includegraphics[width=17cm]{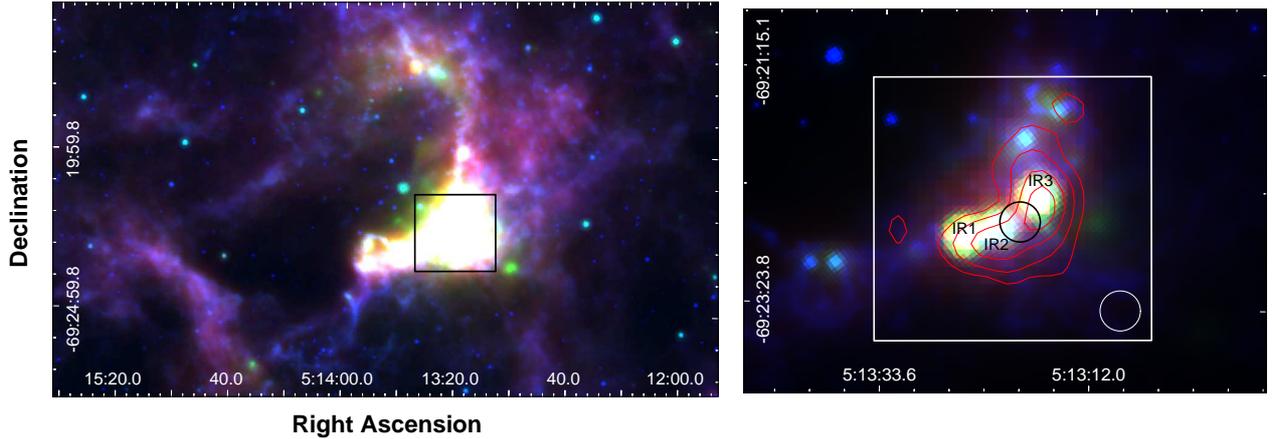}
\caption{Left: Three-colour image where the 8, 24, and 70 $\mu$m emission obtained from the IRAC and MIPS cameras of the Spitzer
Space Telescope are presented in blue, green, and red, respectively. The black box shows the region mapped in the \3 J=3--2 line with
an angular resolution of 22\s. Right: Zoom-in of the mapped region (white box). The colour code is the same as in the left
image but the scales are different. The red contours correspond to the \3 J=3--2 emission integrated between 230 and 245 \ks~with
levels of 2.5, 4.0, 6.0, and 8.0 K \ks. The FWHM beam size of the observations is included in the bottom right corner of the region. The black
circle corresponds to the position where the 16 molecular lines were observed as single pointings.}
\label{present}
\end{figure*}

\section{Observations and data reduction}

The molecular observations were performed between July and August 2010 with the 10 m ASTE telescope \citep{ezawa04}. 
The CATS345 GHz band receiver, a two-single band SIS receiver remotely tunable in the LO frequency range of 324-372 GHz, was used. 
The XF digital spectrometer was set to a bandwidth and spectral resolution of 128 MHz and 125 kHz, respectively.
The spectral velocity resolution was 0.11 \ks~and the half-power beamwidth (HPBW) was 22\s~at 345 GHz. The system temperature
varied from T$_{\rm sys} = 150$ to 250 K and the main beam efficiency was $\eta_{\rm mb} \sim 0.65$.

The data were reduced with NEWSTAR\footnote{Reduction software based on AIPS developed at NRAO,
extended to treat single dish data with a graphical user interface (GUI).} and the spectra processed using the XSpec software 
package\footnote{XSpec is a spectral line reduction package for astronomy which has been developed by Per Bergman at Onsala Space Observatory.}. 
The spectra were Hanning smoothed to improve the signal-to-noise ratio, and in some cases a boxcar smoothing was also applied.
Polynomials between first and third order were used for baseline fitting.

\begin{table}
\caption{Observed molecular lines towards N113.}
\label{obspoints}
\begin{tabular}{lcc}
\hline
\hline
Molecular line            & Integ. time (sec.) & Detection \\
\hline
\2 (3--2)                 &   200              &   yes     \\
\3 (3--2)                 &   1160             &   yes     \\
C$^{18}$O (3--2)          &   1160             &   yes     \\
H$_{2}$D$^{+}$ (1,0--1,1) &   1520             &   no      \\
H$_{2}$CO (5--4)          &   1520             &   no      \\
c-C$_{3}$H$_{2}$ (9--8)   &   1440             &   no      \\
H$_{3}$O$^{+}$ (3,2--2,2) &    480             &   no      \\
HCN (4--3)                &   1520             &   yes     \\
HNC (4--3)                &   800              &   yes     \\
CS (7--6)                 &   600              &   yes     \\
HCO$^{+}$ (4--3)          &   600              &   yes     \\
CH$_{3}$CN (18--17)       &   1280             &   no      \\
CH$_{3}$CN (19--18)       &   1620             &   no(?) \tablefootmark{*}      \\
C$_{2}$H (4--3)           &   1620             &   yes     \\
HNCO (15--14)             &   1280             &   no      \\
DCO$^{+}$ (5--4)          &   1520             &   no      \\     
\hline
\end{tabular}
\tablefoot{
\tablefoottext{*}{see Sect.\,\ref{moleclines}.}}
\end{table}

Several molecular lines in the 345 GHz window were observed towards the center of N113 at RA $=$ 05\hh 13\mm 19.5\ss,
dec.~$=~-$69\d~22\m~37.9\s, J2000 as single pointings (black circle in Fig.\,\ref{present}-right). 
In Table\,\ref{obspoints}, observed molecular lines
and the integration times are listed, indicating whether the detection
was positive or not. Additionally, we mapped a 2\farcm5 $\times$ 2\farcm5 
region centered at RA $=$ 05\hh 13\mm 20\ss,
dec. $= -$69\d 22\m 35.5\s, J2000 in the \3 J=3--2 line (white square in Fig.\,\ref{present}-right). 
This observation was performed in on-the-fly mapping mode achieving an angular sampling of 6\s.

\section{Results and discussion}

\subsection{The molecular cloud}

Figure\,\ref{present} (left) is a three-colour image displaying the mid/far-IR emission in the N113 area where the mapped
region in the \3 J=3--2 line is indicated with a black square. Figure\,\ref{present} (right) displays a zoom-in of the mapped 
region with a different colour scale, which {allows for} the identification of some point-like sources in the IR emission. 
The \3 J=3--2 emission integrated
between 230 and 245 \ks~is presented in contours showing the curved and elongated morphology of the molecular cloud in good agreement
with the IR emission. 
The surveyed area is populated by 18 O and early-B stars (both on the main sequence and evolved), most of them with spectral types 
derived via UBV photometric data \citep{wilcots94}. A few have spectroscopic observations as a mid-OV star
(target s1 in the Wilcots study), sources BI~104 and BI~105 from \citet{massey95}, which are B0.5V and O7V stars, and 
the more widely studied supergiant star, HD269217, that is of type B2[e] (e.g. \citealt{kastner10}).
The main concentration of high-mass YSOs \citep{seale12} and intermediate-mass YSO candidates \citep{carlson12} are located in projection, 
along the CO emission, a configuration commonly found in massive star-forming regions (e.g. \citealt{povich09,book09,chen10}).
The three conspicuous IR sources observed in Fig.\,\ref{present} (right) that appear over the molecular concentration, marked as 
IR1, IR2, and IR3,
are coincident with the compact radio continuum sources detected by \citet{brooks97} and with three YSOs (from southeast to northwest: 
051325.09-692245.1, 051321.43-692241.5, and 051317.69-692225.0) cataloged by \citet{seale12} and \citet{carlson12}. 
As discussed in \citet{wong06} these sources could be young ionizing stars affecting the molecular gas and probably contributing
to the birth of new stars. 

\begin{figure}[h]
\centering
\includegraphics[width=8cm]{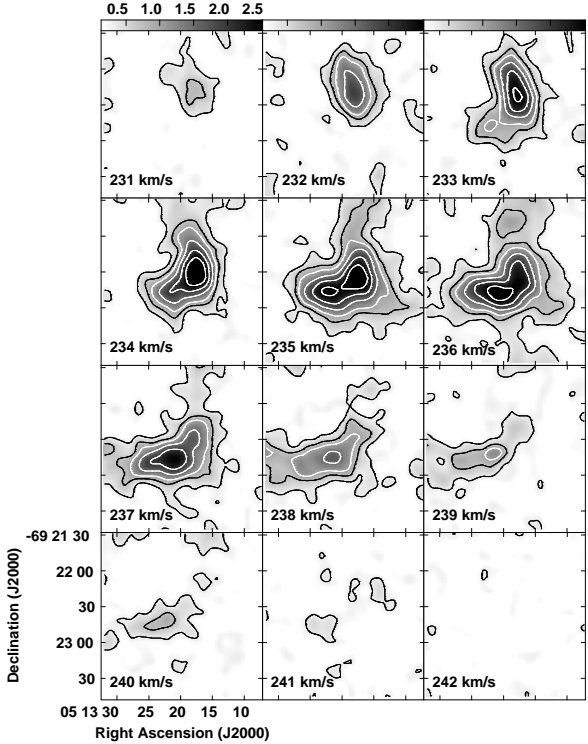}
\caption{Integrated velocity channel maps of the \3 J=3--2 emission every 1 \ks.
The grayscale is displayed at the top of the first panel and is in K \ks, the contour levels are 0.3, 0.7, 1, 2, and 2.5 K \ks.}
\label{panels}
\end{figure}

The morphology and velocity distribution of the molecular gas related to N113 can be seen
in Fig.\,\ref{panels}, where the $^{13}$CO J=3--2 emission is presented in a series of channel maps, 
in the range of 231-242 km s$^{-1}$, integrated in steps of 1 km s$^{-1}$. 
The curved and elongated molecular cloud is resolved into two clumps, one located towards
the northwest and peaking at 233 km s$^{-1}$, the other one slightly eastwards of the center of the surveyed region, 
peaking at 236 km s$^{-1}$.

To derive a rough estimation of the molecular mass,
we assume local thermodynamic equilibrium (LTE). We calculate the excitation temperature from:
\begin{equation}
T_{ex}(3 \rightarrow  2) = \frac{16.59 {\rm K}}{{\rm ln}[1 + 16.59 {\rm K} / (T_{\rm max}(^{12}{\rm CO}) + 0.036 {\rm K})]}
\label{eq1}
\end{equation}
where $T_{\rm max}(^{12}{\rm CO}$) is the $^{12}$CO peak temperature towards the center of the region, obtaining T$_{ex} \sim$ 20 K.
We derive the \2 and \3 optical depths, $\tau_{12}$ and $\tau_{13}$, using (e.g. \cite{scoville86}):
\begin{equation}
\frac{^{12}{\rm T}_{mb}}{^{13}{\rm T}_{mb}} = \frac{1-exp(-\tau_{12})}{1-exp(-\tau_{12}/X)},
\label{eq2}
\end{equation}
where $^{12}$T$_{mb}$ and $^{13}$T$_{mb}$ are the peak temperatures of the \2 and \3 J=3--2 lines at the center of the region, 
and $X = 50$ is the assumed isotope abundance ratio \citep{wang09}. The result is $\tau_{12} \sim 9.5$ and $\tau_{13} \sim 0.2$, which
indicates that the \3 J=3--2 line appears optically thin. Thus, we estimate its column density from:
\begin{equation}
{\rm N} = 8.28 \times 10^{13}~e^{\frac{15.87}{T_{ex}}}\frac{T_{ex}+0.88}{1-e^{\frac{-15.87}{T_{ex}}}} \frac{1}{J(T_{ex})-J(T_{\rm BG})} \int{{\rm T_{mb} ~dv}} 
\label{eq3}
\end{equation}
with
\begin{equation}
J(T) = \frac{h\nu/k}{exp(\frac{h\nu}{kT}) - 1}.
\label{eq4}
\end{equation}
To obtain the molecular hydrogen column density N(H$_{2}$) we assume an abundance ratio of 
[H$_{2}$/\3] $= 1.8 \times 10^{6}$, estimated by \citet{garay02} towards the giant molecular complex No. 37 in the LMC. 
Almost the same value has been obtained for N159W \citep{heik99}, which as N113 is associated with a prominent star-forming region. 
Finally, the mass was derived from:
\begin{equation}
{\rm M} = \mu~m_{{\rm H}} \sum_{i}{\left[ D^{2}~\Omega_{i}~{\rm N_{\it i}(H_{2}}) \right]}, 
\label{eq5}
\end{equation}
where $\Omega$ is the solid angle subtended by the beam size, $D$ is the distance (50~kpc), $m_{\rm H}$ is the hydrogen mass, and $\mu$ 
is the mean molecular weight, assumed to be 2.8 by taking into account a relative helium abundance of 25 \%. 
The summation was performed over all beam positions belonging to the molecular
structure displayed in contours in Fig.\,\ref{present} (right panel). The obtained mass is about $1 \times 10^{4}$ \msol, which is somewhat
lower than the $8.2 \times 10^{4}$ \msol~estimated by \citet{wong06} from the \2 J=1--0 line using a 
standard Galactic CO to H$_{2}$ conversion factor (N$_{\rm H_{2}}$/I$_{\rm CO} = 2.0 \times 10^{20}$ cm$^{-2}$ (K \ks)$^{-1}$).
This discrepancy may be due to that the \3 J=3--2 and the \2 J=1--0 map different extensions of the molecular cloud, and/or
suggests that the Galactic CO to H$_{2}$ conversion factor used by Wong et al. is not appropriate for N113.

From the average of the \3 J=3--2 emission towards the molecular structure 
we obtain an averaged spectrum with $\Delta$v $=$ 5.3 \ks.
Roughly approximating the molecular cloud with a spherical shape of radius 35\s~($R = 8.8$ pc) 
and considering a density profile of $\rho \propto r^{-1}$ we estimate the virial mass from \citep{mac88}:
\begin{equation}
\frac{M_{\rm vir}}{\rm M_{\odot}} = 190 \left( \frac{R}{\rm pc} \right) \left( \frac{\Delta \rm v}{\rm km~s^{-1}} \right)^2, 
\label{eq6}
\end{equation}
which gives $\sim 4.5 \times 10^{4}$ \msol.

\begin{figure}[h]
\centering
\includegraphics[width=8cm]{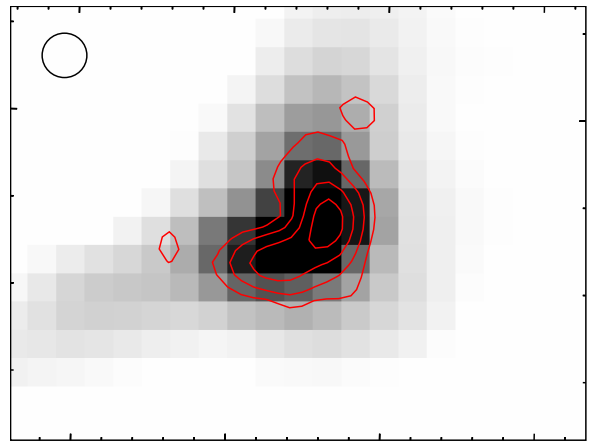}
\caption{500 $\mu$m emission obtained from SPIRE in the Herschel Space Observatory with 
the same \3 J=3--2 contours as shown in Fig.\,\ref{present} (right). The FWHM beam size of the molecular observations 
is included in the top left corner. The angular resolution of the SPIRE image is about 36\s. }
\label{spire}
\end{figure}

Additionally, we estimate the mass of gas of the N113 cloud from the continuum
emission at the far-infrared regime arising from the dust.
To do so we obtain the integrated flux of the continuum
emission at 500~$\mu$m using the calibrated level 2 PLW SPIRE image
extracted from the Herschel Data Archive (ObsID:1342202224).
Figure \ref{spire} displays the 500~$\mu$m emission with contours of the
\3 J=3--2, showing that both emissions are very similar in morphology and size.
From the radiative transfer equation and assuming an optically thin
medium we can estimate the gas mass in the cloud as:
\begin{equation}
M_{gas}=\frac{S_{\lambda} D^2}{\kappa_d(\lambda) x_{d}  B_{\lambda}(T_{d})},
\label{eq7}
\end{equation}
where $S_{\lambda}$, $D$, $\kappa_d(\lambda)$, $x_{d}$, and
$B_{\lambda}(T_{d})$ are the integrated flux at 500~$\mu$m, the
distance, the dust absorption coefficient, the dust-to-gas mass ratio, 
and the Planck's law at the $T_d$ dust temperature, respectively. 
Considering the obtained integrated flux at 500~$\mu$m of
about 50~Jy, a distance of 50~kpc, $\kappa_d(500\mu m) = 1.14$ for the LMC \citep{weidraine}, a $T_d$ of about 24~K
and $x_{d} = 1.7 \times 10^{-3}$ \citep{verdugo11} we derive a
mass of gas of about $7 \times 10^{3}$~M$_{\odot}$. 
This value is almost two orders of magnitude lower than the mass estimated by \citet{wang09} from 
the 1.2 mm dust continuum. The $M_{vir}/M_{gas}$ ratio is larger than unity, as found in several clouds in 
N11 \citep{herrera}, a bright HII region in the LMC hosting several star clusters.

\subsection{Molecular lines}
\label{moleclines}

\begin{figure}[h]
\centering
\includegraphics[width=5cm]{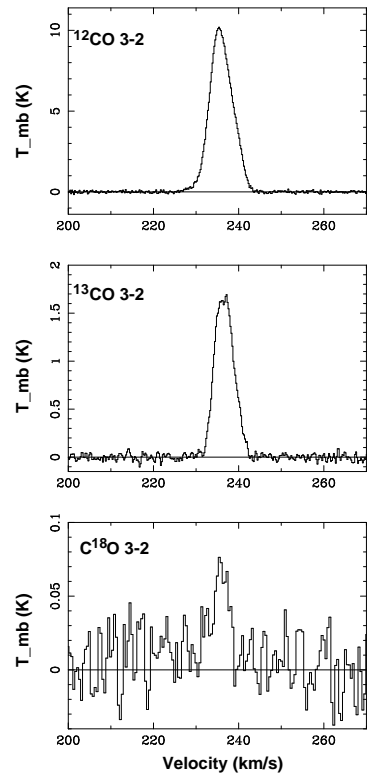} 
\caption{CO isotopologues  
obtained towards the center of N113. The rms noise levels of each spectrum are: 80, 35, and 30 mK, respectively, 
and the displayed channel spacings are about 0.2 \ks~for the \2 and \3, and 0.4 \ks~for the C$^{18}$O.}
\label{spectra1}
\end{figure}

\begin{figure}[h]
\centering
\includegraphics[width=9cm]{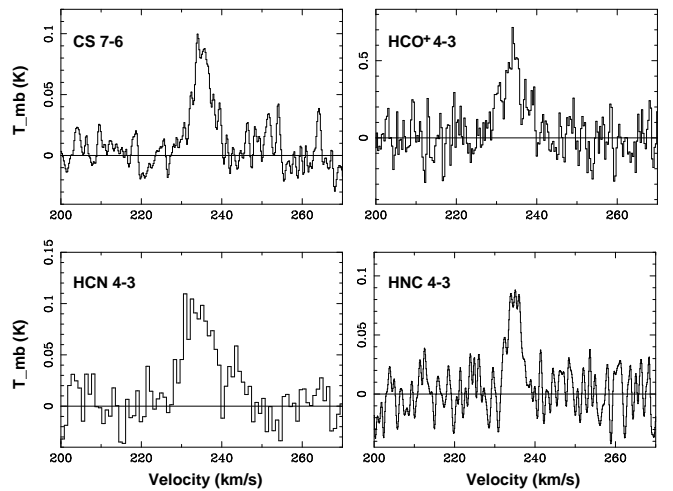}
\caption{Spectra of CS J=7--6 and \H, HCN, and HNC J=4--3 
obtained towards the center of N113. The rms noise levels of each spectrum are: 14, 90, 20, and 13 mK, respectively,
and the displayed channel spacings are about 0.2 \ks~for the CS, \H, and HNC, and 1.2 \ks~for the HCN.}
\label{spectra2}
\end{figure}

In Figs.\,\ref{spectra1} and\,\ref{spectra2} we present the spectra of the molecular lines successfully observed towards the 
position indicated with a black circle in Fig.\,\ref{present} (right). 
Figure\,\ref{spectra1} shows the CO isotopologues, and Fig.\,\ref{spectra2}, the emission of the detected rarer lines.
The line parameters given in Table\,\ref{lines} were obtained from single-component Gaussian fits. 
All the spectra have signals well above the 3$\sigma$, except
the C$^{18}$O J=3--2 line, for which the signal is evident but the noise is high. It is important to note that the HNC and
HCN lines are detected for the first time towards N113 in the J=4--3 transition. On the other side, the CS J=7--6 line was 
tentatively detected by \citet{wang09} and in this study we confirm its detection. Our CS central velocity and $\Delta$v 
are in agreement, within the errors, with the previous tentative detection. Taking into account that the critical density
of the HCN J=4--3 line is about $10^{8}$ cm$^{-3}$ \citep{taka07}, we conclude that we are observing a high density 
molecular clump, which is in agreement with the conspicuity of the CS J=7--6 line, also a tracer of high density gas.

\begin{table}
\caption{Line parameters for the molecular lines presented in Figs.\,\ref{spectra1} and \ref{spectra2}.}
\label{lines}
\tiny
\begin{tabular}{lcccc}
\hline
\hline
Molecular line            & $\int{\rm T_{mb} dv}$  & T$_{\rm mb}$ peak & v$_{\rm LSR}$    & $\Delta$v (FWHM)  \\
                          &   (K \ks)              &   (K)             &   (\ks)         &  (\ks)            \\ 
\hline 
\noalign{\smallskip}
\2 (3--2)                 &   66.20$\pm$2.20         &   9.90$\pm$0.30     &  235.84$\pm$0.10  &  6.25$\pm$0.20         \\
\3 (3--2)                 &   9.52$\pm$1.50          &   1.73$\pm$0.32     &  236.58$\pm$0.47  &  5.23$\pm$1.10         \\
C$^{18}$O (3--2)          &   0.35$\pm$0.10          &   0.05$\pm$0.02     &  235.68$\pm$0.55  &  4.73$\pm$1.70         \\
CS (7--6)                 &   0.54$\pm$0.18          &   0.08$\pm$0.02     &  235.15$\pm$0.80  &  5.75$\pm$1.20        \\
HCO$^{+}$ (4--3)          &   3.40$\pm$0.20          &   0.47$\pm$0.10     &  233.90$\pm$0.95  &  7.15$\pm$1.50        \\
HCN (4--3)                &   0.70$\pm$0.15          &   0.09$\pm$0.02     &  234.28$\pm$0.75  &  8.22$\pm$1.50        \\
HNC (4--3)                &   0.35$\pm$0.10          &   0.08$\pm$0.02     &  234.87$\pm$0.70  &  3.85$\pm$1.00       \\
\hline
\end{tabular}
\end{table}

\begin{figure}[h]
\centering
\includegraphics[width=8cm]{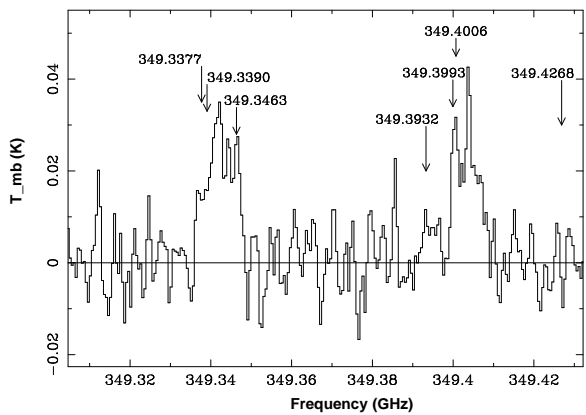}
\caption{Spectrum in the frequency range 349.30 -- 349.43 GHz. The arrows indicate the frequencies where molecular 
emission lines are expected, according to the NIST data base (see Table\,\ref{c2h+}). The rms noise is about 6 mK
and the displayed channel spacing is about 0.5 MHz.}
\label{c2h}
\end{figure}

Figure\,\ref{c2h} shows the obtained spectrum in the frequency range 349.30 -- 349.43 GHz, which is populated
by several lines and fine structure components of C$_{2}$H and CH$_{3}$CN. The frequencies where molecular emission lines are 
expected according to the NIST data base\footnote{http://www.nist.gov/pml/data/micro/index.cfm}, 
are indicated in the figure and listed in Table\,\ref{c2h+}.
In the spectrum presented in Fig.\,\ref{c2h} two peaks can be distinguished, likely due to the C$_{2}$H (4--3) fine structure 
transitions. The peak centred around 349.342 GHz is composed 
by two C$_{2}$H blended hyperfine lines and probably by one line of the CH$_{3}$CN 19--18 transition. 
The other peak, centred around 349.403 GHz, is also composed by two C$_{2}$H blended hyperfine lines. 
\citet{chin97} and \citet{wang09} have reported the detection of C$_{2}$H (1--0) {\it J}=3/2--1/2 F=2-1 and 1-0 lines towards N113 
and in this study we are presenting the first detection of C$_{2}$H (4--3) towards this region.
Regarding the CH$_{3}$CN 19(4)--18(4) line, probably blended in the peak centred around 349.345 GHz, it is not possible to confirm 
its detection in our observations. Besides, the non-detection of the CH$_{3}$CN 18--17 line, a lower transition, also suggests a
non-detection for the J=19--18 transition.

\begin{table}
\caption{Molecular lines within the 349.30 -- 349.43 GHz range (see spectrum in Fig.\,\ref{c2h}).}
\label{c2h+}
\small
\begin{tabular}{ccc}
\hline
\hline
Frequency & Molecule  &   Transition (quantum numbers) \\
(GHz)     &           &                                \\
\hline
\noalign{\smallskip}
349.3377  &  C$_{2}$H         & 4--3 {\it J}=9/2-7/2 F=5-4        \\
349.3390  &  C$_{2}$H         & 4--3 {\it J}=9/2-7/2 F=4-3        \\
349.3463  &  CH$_{3}$CN       & 19(4)--18(4)                   \\
349.3932  &  CH$_{3}$CN       & 19(3)--18(3)                 \\
349.3993  &  C$_{2}$H         & 4--3 {\it J}=7/2-5/2 F=4-3        \\
349.4006  &  C$_{2}$H         & 4--3 {\it J}=7/2-5/2 F=3-2        \\
349.4268  &  CH$_{3}$CN       & 19(2)--18(2)        \\
\hline
\end{tabular}
\end{table}

\citet{beuther08} observed C$_{2}$H (4--3) towards a Galactic sample 
of star forming regions in different evolutionary stages, including infrared dust clouds (IRDCs), high mass protostellar objects (HMPOs),
and ultracompact HII regions (UCHIIs). They found that the C$_{2}$H lines are detected independently of the evolutionary stage
of the sources, but the UCHIIs regions exhibit line widths in both C$_{2}$H (4-3) main peaks significantly broader 
than the others objects (on average about 5.5 \ks).
Assuming that the peaks in the spectrum of Fig.\,\ref{c2h} are due exclusively to emission of C$_{2}$H,
and converting the frequency into velocity, we measure the FWHM of both peaks as 6.8 and 6.4 \ks, respectively,
indicating that both are broad lines as those measured by \citet{beuther08} towards UCHIIs.
This result is in agreement with the presence of the compact radio continuum sources detected by \citet{brooks97} in N113. 
Besides, the position where the molecular lines were observed in this study lies between the two most intense continuum sources in 
the study of \citet{wong06} (sources 2 and 3; IR2 and IR3 in Fig.\,\ref{present}-right). 
The authors have estimated that the equivalent to one or two O6 stars is needed 
to produce the measured fluxes in the radio continuum (at 24 and 86 GHz) towards source 2, which,
as discussed in their study, should be young ionizing stars affecting the molecular gas.
Indeed, the C$_{2}$H can be formed and/or replenished after destruction in earlier stages, in PDRs, through 
C$_{2}$H$_{2} + h\nu \rightarrow$ C$_{2}$H $+$ H. The neutral-neutral
reaction CH$_{2}$ $+$ C $\rightarrow$ C$_{2}$H $+$ H can also produce C$_{2}$H,
where the precursor carbon atom is formed through the photodissociation of CO (\citealt{miett14}, and references therein).
Therefore, we suggest that the chemistry involving this radical in N113 can be similar to that proceeding in Galactic PDRs.

In Table\,\ref{ratios} the observed integrated intensity ratios are listed for the lines presented in Table\,\ref{lines},
together with a comparison with ratios obtained using the J=1--0 transitions from \citet{chin97} and from \citet{wang09} for the 
transitions indicated in the Table. Our results are in good agreement with the ratios obtained from lower transitions, except 
for the \H/HCN and HNC/\H~ratios.
The \H/HCN ratio is larger than unity as was found in several Magellanic giant molecular clouds, suggesting that
the ion abundance is higher than in Galactic clouds, where this ratio was found to be lower than unity \citep{stacey91,chin97}. 
This must be due to higher UV fields in the Magellanic environments.
Additionally, the low nitrogen abundance in the LMC (e.g. \citealt{hunter07,bekki10}) can also contribute in the increment of the \H/HCN ratio.
Furthermore, it seems that the \H/HCN ratio increases for increasing rotational transitions, which suggests that the physical conditions
in N113 may favour the excitation of the \H~higher transitions more efficiently than those of HCN. 
If the \H~and HCN emission occur in the same region, the increment in the \H/HCN ratio with the rotational transitions may reflect different 
critical densities, with HCN being selectively de-excited at higher transitions. Indeed, even though the $E_{u}/k_{B}$ factor is similar in both 
molecular species, their critical densities vary between
different J$_{u}$--J$_{l}$ transitions (n$_{crit}$(HCN)/n$_{crit}$(\H) $\sim$ 5--7 for J $=$ 1--0, 3--2, and 4--3; \citealt{papa07}).
Similar cases of this \H/HCN ratio behaviour were found towards the Galactic NGC 1333-IRAS 2A outflow \citep{jorge04} and towards the
nuclear region of M82 \citep{seaquist00}. 
Regarding the HCN/HNC ratio, it is larger than unity in the J=4--3 line, as is the case for the lower transitions,
supporting the prediction by \citet{chin97}
that the ratio would be large in warm gas subject to strong UV heating, but approaches unity in cloud cores. Also \citet{wang09} point
out that the HCN/HNC ratio larger than unity may indicate a PDR scenario, which is in agreement with the C$_{2}$H chemistry discussed above.

\begin{table}
\caption{Integrated intensity ratios towards N113.}
\label{ratios}
\small
\begin{tabular}{lccc}
\hline
\hline
Ratio                                 &  this work    & \citet{chin97}    & \citet{wang09}  \\
                                      &               &  (all J=1--0)     &                 \\
\hline
\noalign{\smallskip}
$\frac{\rm ^{12}CO}{\rm ^{13}CO}$     &  $6.9\pm1.1$  &  $7.28\pm0.70$  & $4.50\pm0.16$ {\tiny (J=3--2)}     \\
\noalign{\medskip}
$\frac{\rm ^{13}CO}{\rm C^{18}O}$     &  $27.0\pm8.8$\tablefootmark{*}  &  $35.8\pm3.1$   & $39.8\pm2.7$ {\tiny (J=1--0)}       \\
\noalign{\medskip}
$\frac{\rm HCO^+}{\rm HCN}$           &  $4.8\pm1.1$  &  $1.34\pm0.06$  & $1.47\pm0.10$ {\tiny (J=1--0)}   \\
                                      &               &                 & $2.15\pm0.47$ {\tiny (J=3--2)} \\
\noalign{\medskip}
$\frac{\rm HCN}{\rm HNC}$             &  $2.0\pm0.7$  &  $2.8\pm0.2$    & $2.6\pm0.2$   {\tiny (J=1--0)}     \\
\noalign{\medskip}
$\frac{\rm HNC}{\rm HCO^{+}}$         &  $0.10\pm0.03$ & $0.26\pm0.01$  & $0.26\pm0.01$ {\tiny (J=1--0)}     \\
\noalign{\smallskip}
\hline
\end{tabular}
\tablefoot{
\tablefoottext{*}{Tentatively ratio due to the high noise in the C$^{18}$O spectrum.}}
\end{table}

\begin{figure}[h]
\centering
\includegraphics[width=6cm]{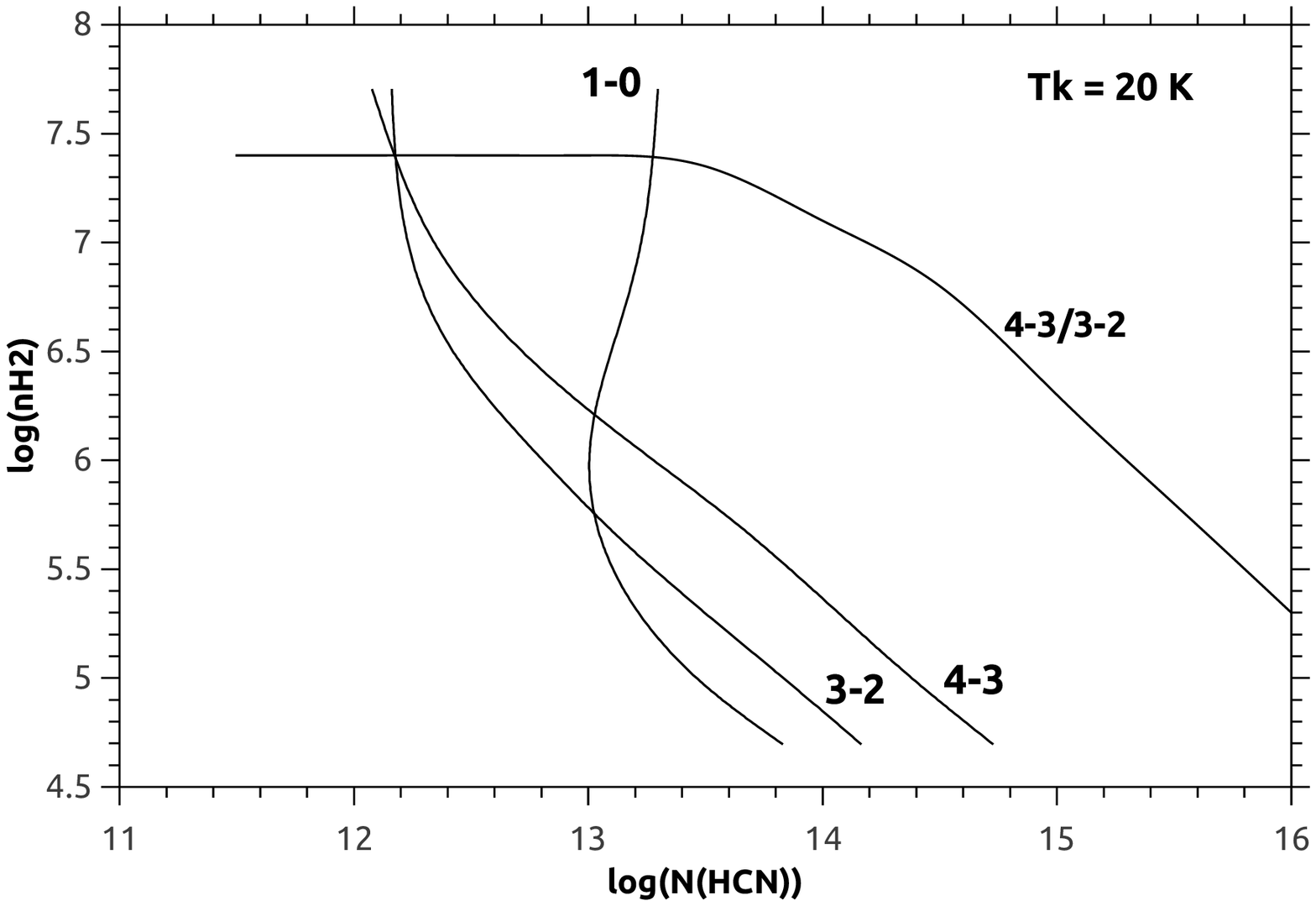}
\includegraphics[width=6cm]{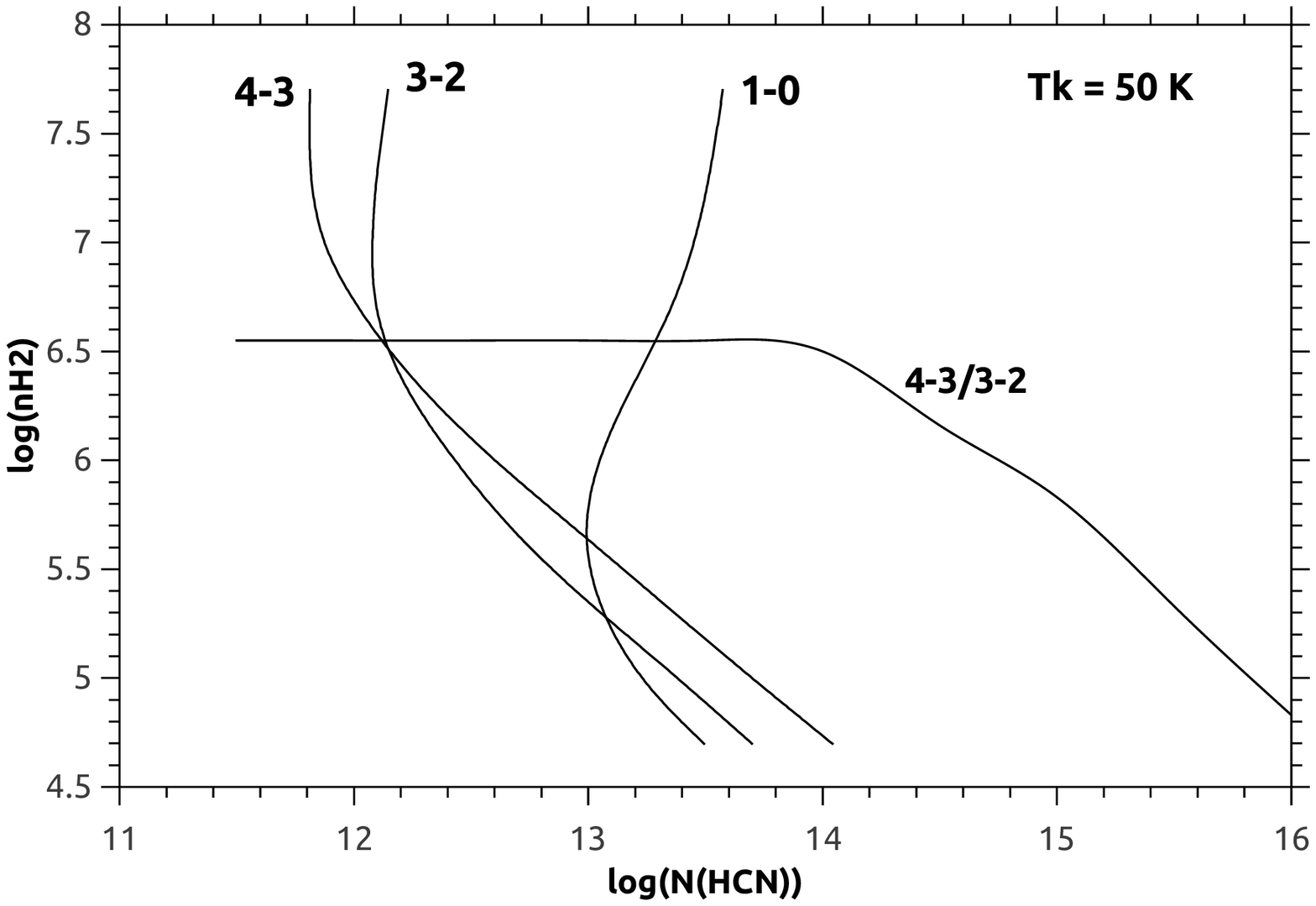}
\includegraphics[width=6cm]{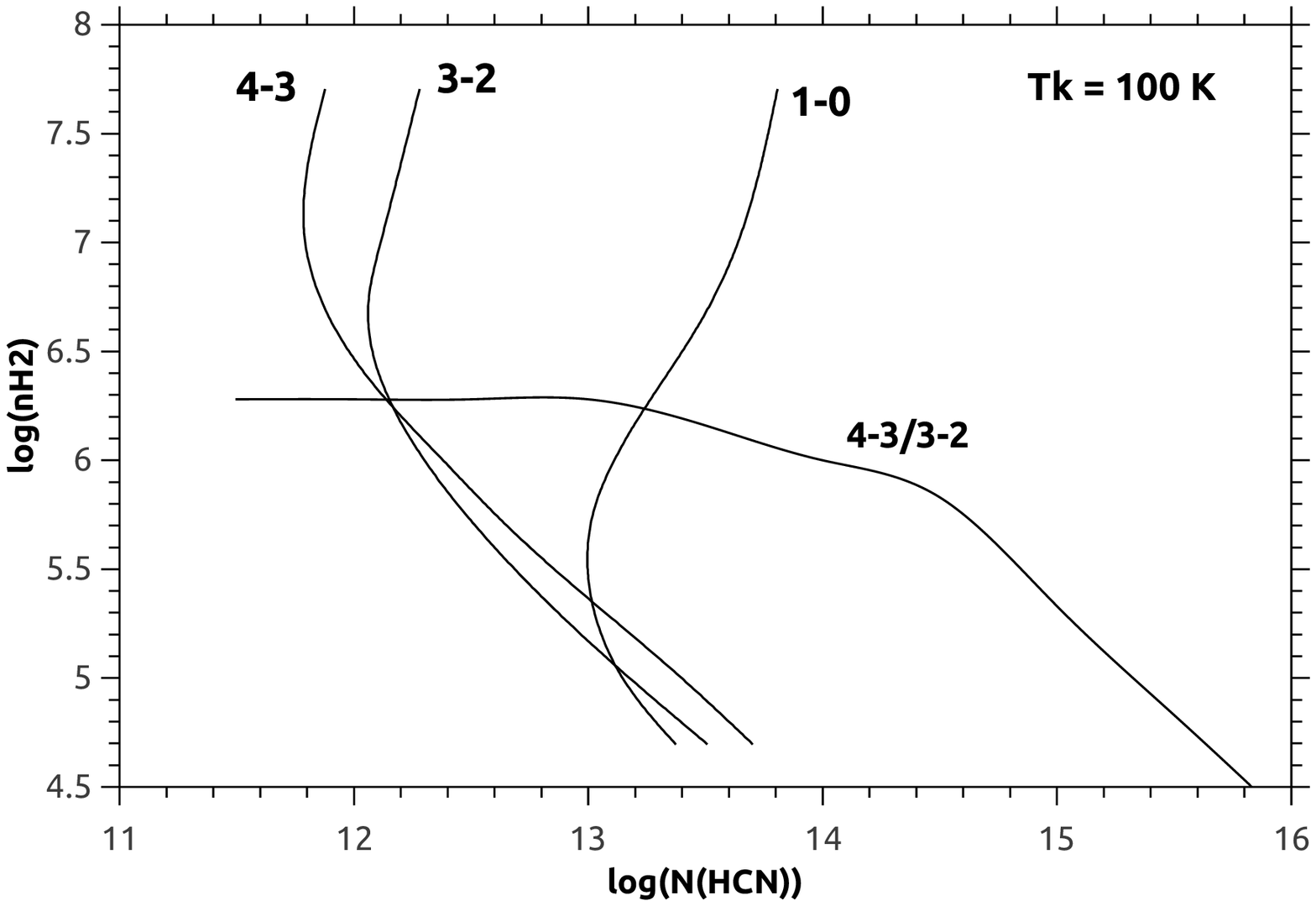}
\caption{Radex calculations for the HCN J=4--3 and 3--2 lines for kinetic temperatures of 20, 50, and 100 K. The
results for HCN J=1--0 are included for comparison with \citet{wang09}. }
\label{radex}
\end{figure}

Using the derived HCN J=4--3 parameters listed in Table\,\ref{lines} and the HCN J=3--2 and J=1--0 parameters presented in \citet{wang09} we
perform a non-LTE study of this molecular species using RADEX \citep{vander06}.
The RADEX model uses the mean escape probability approximation for the radiative transfer equation.
As done by \citet{wang09}, we correct for beam dilution by calculating T$_{\rm mb}^{\prime} =$ T$_{\rm mb}$/$\eta_{bf}$,
with $\eta_{bf} = \theta_{s}^{2}/(\theta_{s}^{2} + \theta_{b}^{2})$, where $\theta_{b}$ and $\theta_{s}$ are the beam and 
the source size. Following Wang et al., a source size of 40\s~was assumed. Then we run the RADEX code using the measured $\Delta$v 
to fit T$_{\rm mb}^{\prime}$ for each transition.
Figure\,\ref{radex} shows the RADEX calculations for the HCN, displaying the expected H$_{2}$ density and N(HCN) pairs that 
correspond to a given T$_{\rm mb}^{\prime}$ and 4--3/3--2 intensity ratio.
The calculations were made for kinetic temperatures of 20, 50, and 100 K,
as done in previous studies (see \citealt{wang09} and references therein). The obtained results from the J=4--3 and J=3--2 lines are presented 
in Table\,\ref{tradex}. The results from the J=3--2 and J=1--0 lines are similar to those obtained by \citet{wang09},
that is a density of several $10^{5}$ cm$^{-3}$, and N(HCN) of about $1 \times 10^{13}$ cm$^{-2}$. Thus
we conclude that the HCN column density ranges from 1.4 $\times 10^{12}$ to about 1 $\times 10^{13}$ cm$^{-2}$, 
and the density varies from some 10$^{5}$ to a few 10$^{7}$ cm$^{-3}$. Our results confirm that the HCN emission emanates from a very high
density region in N113. The density of this region ranges between the \H~and HCN J=4--3 critical densities,
which can explain the dependency of the \H/HCN ratio on the rotational transitions discussed above.

\begin{table}
\caption{Radex results from the HCN 4--3 and 3--2 lines.}
\label{tradex}
\centering
\begin{tabular}{lcccc}
\hline
\hline
T$_{k}$ (K)            & n$_{\rm H_{2}}$ (cm$^{-3}$) & N(HCN) (cm$^{-2}$) & $\tau_{4-3}$ & $\tau_{3-2}$  \\
\hline
\noalign{\smallskip}
20             &   2.51 $\times 10^{7}$            &   1.47 $\times 10^{12}$  & 0.014 & 0.018     \\
50             &   3.23 $\times 10^{6}$            &   1.38 $\times 10^{12}$  & 0.014 & 0.019    \\
100            &   1.82 $\times 10^{6}$          &   1.43 $\times 10^{12}$  & 0.014 & 0.020    \\
\hline
\end{tabular}
\end{table}

\section{Summary}

N113 is an HII region located in the central part of the Large Magellanic Cloud with an associated
molecular cloud very rich in molecular species. At present, most of the observed molecular lines cover the frequency
range 85--270 GHz, requiring a survey at higher frequencies in the 345 GHz window in order to have a
more complete understanding of the chemistry and excitation conditions of the region. To perform that, we mapped 
a region of 2\farcm5 $\times$ 2\farcm5~centered at N113 using the Atacama Submillimeter Telescope Experiment 
in the \3 J=3--2 line with an angular and spectral resolution of 22\s~and 0.11 \ks, respectively, and observed 
16 molecular lines as single pointings towards its center. The main results can be summarized as follows:

(1) The N113 associated molecular cloud mapped in the \3 J=3--2 line shows a curved and elongated morphology in good agreement
with the IR emission. From this line we estimated LTE and virial masses for the molecular cloud of about 
$1\times10^{4}$~and $4.5\times10^{4}$ \msol, 
respectively. Additionally, from the dust continuum emission at 500 $\mu$m we obtained a mass
of gas of $7 \times 10^3$ \msol.
 
(2) Towards the center of the N113 molecular cloud we detected emission from: 
\2, \3, C$^{18}$O (3--2), HCN, HNC, \H, C$_{2}$H (4--3), and CS (7--6),
being the first reported detection of HCN, HNC, and C$_{2}$H in the J=4--3 line from this region.
CS (7--6), which was previously tentatively detected, was confirmed in this study. The detection of HCN (4--3) and
CS (7--6) reveals a very high density region .

(3) The observed C$_{2}$H (4--3) presents two peaks due to its fine structure transitions. 
We suggest that the chemistry involving the C$_{2}$H in N113 is similar to that in Galactic PDRs.

(4) We found that the HCN/HNC ratio is larger than unity in the J=4--3 line, as is the case for lower transitions,
supporting the prediction that this ratio would be large in warm gas subject to strong UV heating, indicating a PDR scenario, 
which is in agreement with our finding of broad C$_{2}$H lines. 
Additionally, we found that \H/HCN ratio increases with increasing rotational transitions, showing different
critical densities for both molecular species.

(5) Using the parameters derived from our HCN J=4--3 observation and previous results from HCN J=3--2 and J=1--0 we
performed a non-LTE study of this molecule. Our results confirm that the HCN emission emanates from a very high
density region in N113, with densities ranging between some 10$^{5}$ and 10$^{7}$ cm$^{-3}$.

\begin{acknowledgements}
We thank the anonymous referee for her/his helpful comments and corrections.
The ASTE project is led by Nobeyama Radio Observatory (NRO), a branch
of National Astronomical Observatory of Japan (NAOJ), in collaboration
with University of Chile, and Japanese institutes including University of
Tokyo, Nagoya University, Osaka Prefecture University, Ibaraki University,
Hokkaido University, and the Joetsu University of Education.
S.P. and M.O. are members of the {\sl Carrera del 
investigador cient\'\i fico} of CONICET, Argentina. 
This work was partially supported by grants awarded by CONICET, ANPCYT and UBA (UBACyT) from Argentina.
M.R. wishes to acknowledge support from CONICYT through FONDECYT grant No 1140839.

\end{acknowledgements}

%%%%%%%%%%%%%%%%%%%%%%%%%%%%%%%%%%%%%%%%%%%%%%%%%%%%%%%%%%%%%%%%%%%%%
\bibliographystyle{aa} % style aa.bst
\bibliography{ref} % your references Yourfile.bib

\end{document}